\documentclass[amssymb,prb,twocolumn,superscriptaddress,floats,showkeys,showpacs]{revtex4-1}
\usepackage[T1]{fontenc}
\usepackage[caption=false]{subfig}
\usepackage{bm}
\usepackage{graphicx}
\usepackage{amssymb}
\usepackage{upgreek}
\usepackage{natbib}
\usepackage{amsfonts}
\usepackage{amsmath} 
\usepackage{hyperref} 
\hypersetup{colorlinks,citecolor=blue, filecolor=blue ,linkcolor=blue , urlcolor=blue, pdftex}

\usepackage{tabularx}

\begin{document}

\newcommand{\HfS}{$\text{HfS}_{2}$ }

\title{Pressure-driven phase transitions in bulk \HfS}


\def \FUW{Institute of Experimental Physics, Faculty of Physics, University
of Warsaw, Pasteura 5, 02-093 Warsaw, Poland}
\def \China{Hefei Innovation Research Institute, School of Microelectronics, Beihang University, Hefei 230013, P. R. China}
\def \CENT{Centre of New Technologies, University of Warsaw, Banacha 2c, 02-097 Warsaw, Poland}
\def \Wroclaw{Department of Semiconductor Materials Engineering, Faculty of Fundamental Problems of Technology, Wroc\l{}aw University of Science and Technology, Wybrze\.ze Wyspia\'nskiego 27, 50-370, Wroc\l{}aw, Poland}
\def \Spain{Geosciences Barcelona (GEO3BCN), CSIC, Llu\'is Sol\'e i Sabar\'is s.n., 08028, Barcelona, Catalonia, Spain}

\author{M. Grzeszczyk} \affiliation{\FUW}
\author{J. Gawraczy\'nski}\affiliation{\CENT}
\author{T. Wo\'zniak}\affiliation{\Wroclaw}
\author{J. Ib{\'a}{\~n}ez}\affiliation{\Spain}
\author{Z. Muhammad}\affiliation{\China}
\author{W. Zhao}\affiliation{\China}
\author{M. R. Molas}\affiliation{\FUW}
\author{A. Babi\'nski}\affiliation{\FUW}

\begin{abstract} 
The effect of hydrostatic pressure up to 27 GPa on the Raman scattering (RS) in bulk \HfS is investigated. 
There are two transformations of RS spectra, which take place during compression at pressure between 5.7 GPa and 9.8 GPa as well as between 12.8 GPa and 15.2 GPa. 
Seven vibrational modes can be observed after the transformation, as compared to four modes before the transformation. The observed change suggests structural change in the material of yet unknown nature. The frequencies of the RS modes observed above the transformation change linearly with pressure and corresponding pressure coefficients have been determined. 
The other transition manifests itself as a change in the RS lineshape.
While a series of well-defined RS modes are observed under pressure below the transition, broad spectral bands can be seen at higher pressure. 
The overall lineshape of the spectra resembles that of disordered materials. 
The lineshape does not change during decompression, which suggests permanent nature of the high-pressure transition.  
\end{abstract}

\maketitle

\section{Introduction \label{sec:Intro}}

Layered transition metal dichalcogenides (TMDs) and in particular their few-layer structures have been drawing the attention of researchers for more than a decade now. 
While about 60 layered TMDs are recognized, until now, the attention of researchers is focused on Mo- and W-based compounds. Although basic properties of several TMDs in their bulk form are known for years \cite{wilson1969}, there is still room for studies of $e.g.$ their properties under hydrostatic pressure.

In this work we address the effect of hydrostatic pressure on hafnium disulphide (\HfS), a member of group IVB TMD.  
\HfS has been recently shown to have very effective electrical response \cite{xu2015ultrasensitive, kanazawa2016transistor}, which justifies a need to uncover basic properties of the material.
The evolution of lattice dynamics of \HfS with hydrostatic pressure is studied.
We follow RS spectra as a function of pressure to investigate possible structural transformations of the material as a first-order phase transition has been recently reported in Ref.\cite{ibanez2018high}. 

\section{Samples and experimental setups \label{sec:Methods}}

\begin{figure}[t]
	\includegraphics[width=.5\textwidth]{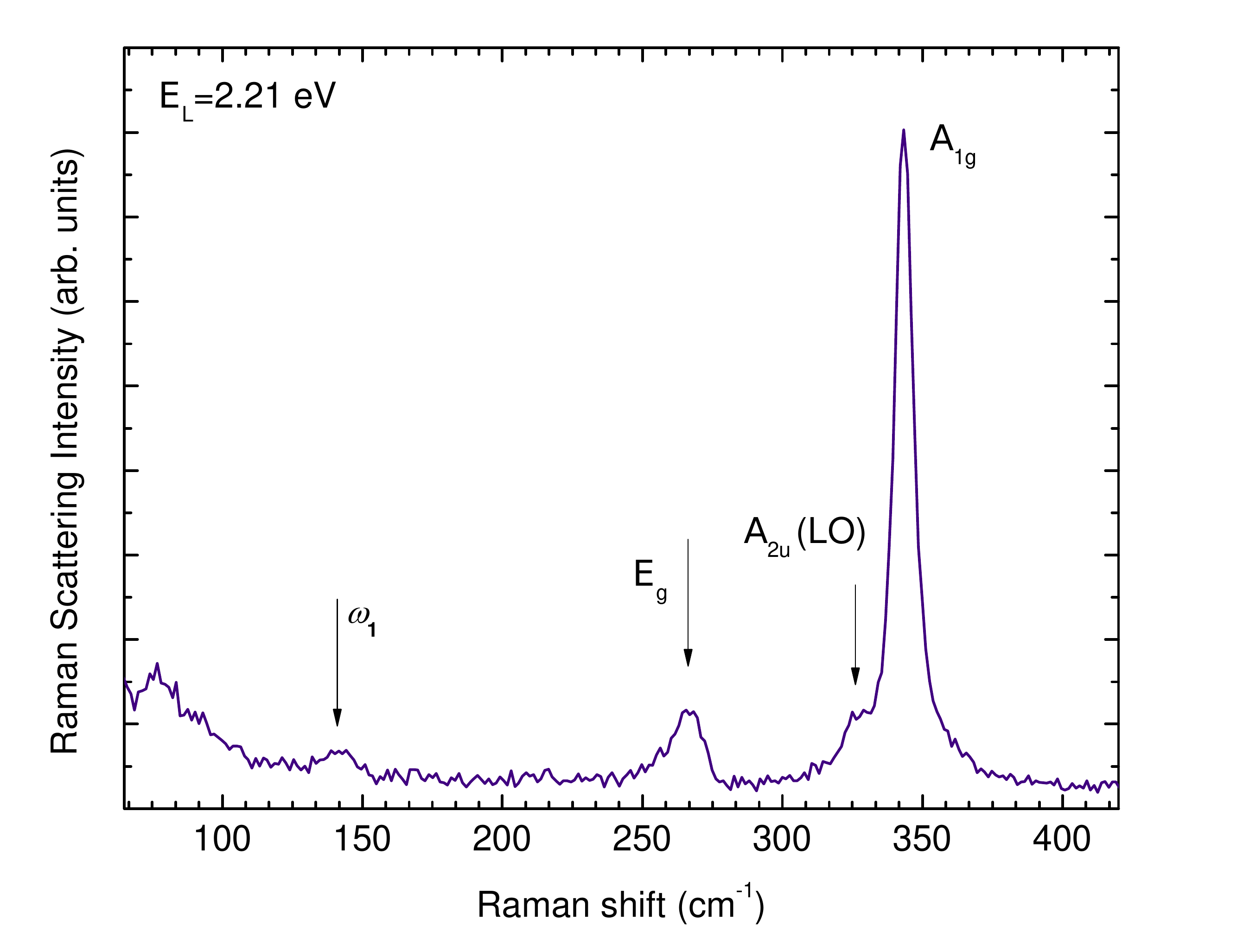}
\caption{Raman scattering spectrum of bulk \HfS at ambient pressure.} \label{fig:0}
\end{figure}

\begin{figure*}[t!]
	\includegraphics[width=\textwidth]{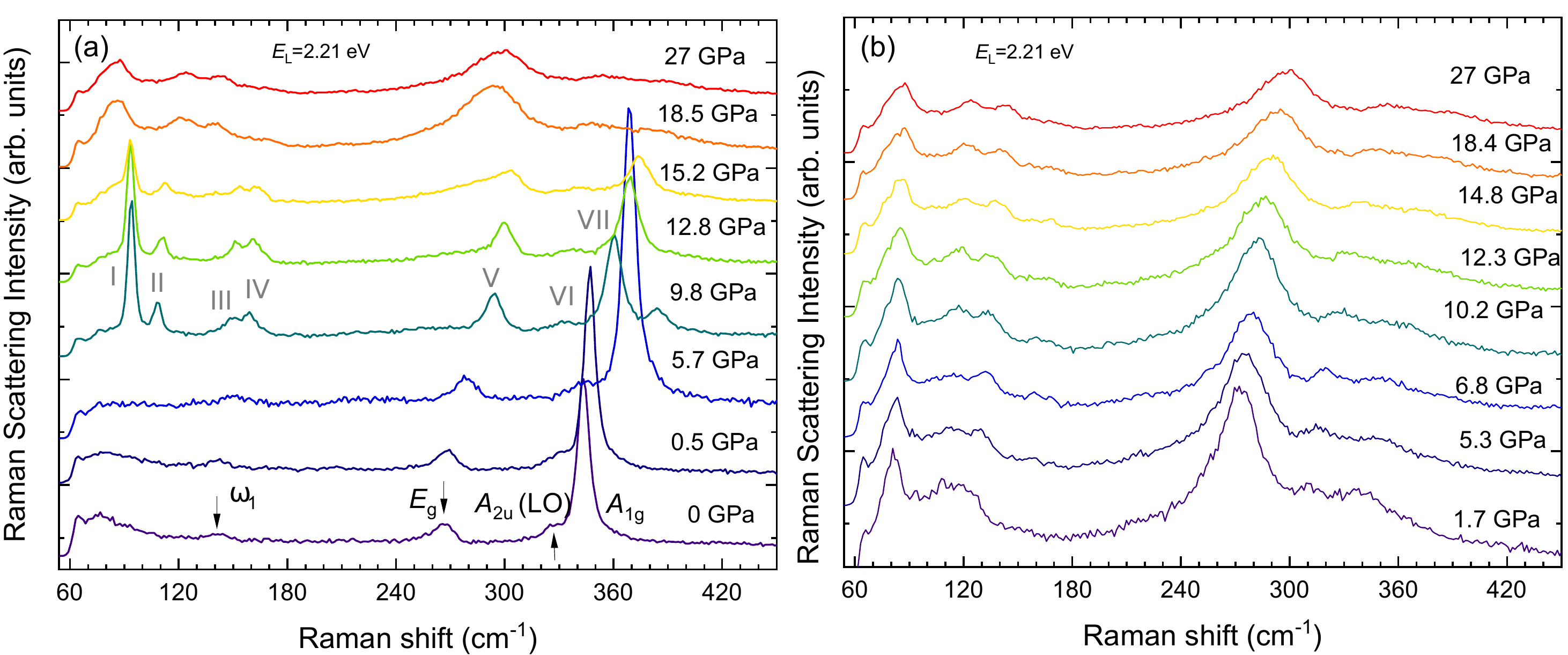}
\caption{Raman scattering spectra of  bulk \HfS  as a function of hydrostatic pressure for compression (a) and decompression (b) cycle} \label{fig:1} 
\end{figure*}

Single crystals of \HfS have been synthesized in two-zone chemical vapour transport (CVT) furnace. 
First, precursor materials were sealed in quartz tubes 25 cm long and of diameter of 8 cm. 
Iodine was used as a transport agent. 
The reaction and growth temperatures were set to 1050 K and 950 K, respectively.
The growth was performed continuously for 120 hours. 
With higher stability of \HfS against iodine, CVT growth first starts at elevated temperatures. 
In the vapour phase HfI and HfI2 can also dominate the direction of reaction. 
After 120 hours the reaction was stopped automatically and the reactor was cooled down to room temperature in 5 hours. 
High quality \HfS single crystals of 0.8-1 cm$^2$ size were grown in the low-temperature part of the reactor. 
The pattern of powder X-Ray diffraction (XRD) of the investigated material was found to be in perfect agreement with that of the of the octahedral 1T phase of \HfS \cite{lukovsky1973IR}. 
A very minor amount of an unidentified phase was also detected in the crystal by XRD technique \cite{ibanezPrivate}.

The high-pressure RS measurements were conducted at room temperature in a nonmagnetic diamond anvil cell (DAC) made of BeCu alloy. 
Crystallographic oil was adopted as the pressure transmitting medium. The fluorescence line of ruby was used to calibrate pressure. 
RS spectra were excited using \mbox{$\lambda$=561~nm} (2.21~eV) radiation from a diode-pumped laser. The excitation light was focused by means of a 50x long-working distance objective with a 0.42 numerical aperture (NA) producing a spot of about 1~$\upmu$m diameter. The signal was collected via the same microscope objective, sent through 0.75~m monochromator and then detected by using a liquid nitrogen-cooled charge-coupled device (CCD) camera. 
The spectral resolution of the setup in the studied energy range of the RS experiment was equal to 0.7~cm$^{-1}$.

\section{Results \label{sec:Results}}

There are six normal vibrational modes at the $\Gamma$ point of 1T-\HfS \cite{lukovsky1973IR}: \\
\vspace{-35pt}
\begin{center}
\begin{equation*}
     \Gamma=A_{1g}+E_g+2A_{2u}+2E_u
\end{equation*}
\end{center}

\noindent The A$_{1g}$ and E$_g$ modes are Raman-active, the infrared-active modes A$_{2u}$ and E$_u$ are split into LO and TO branches (for the atom displacement pattern see $e.g.$ \cite{neal2021Raman}), and other A$_{2u}$ and E$_u$ modes are acoustic\cite{lukovsky1973IR} . 

The measured RS spectrum of bulk \HfS at ambient pressure consists of four modes (see Fig.~\ref{fig:0}), the frequencies of which
equal approx. to: 141~cm$^{-1}$, 266~cm$^{-1}$, 325~cm$^{-1}$, and 344~cm$^{-1}$.
This is consistent with previous studies \cite{ibanez2018high,roubi1988resonance,cingolani1987raman}. There is an agreement in literature on the attribution of peaks at 266 cm$^{-1}$ and 344 cm$^{-1}$ to in-plane E$_g$ and out-of-plane A$_{1g}$ Raman-active modes, respectively. The peak at 325 cm$^{-1}$ is attributed to infrared-active E$_u$ (LO) \cite{cingolani1987raman,neal2021Raman} or A$_{2u}$(LO) \cite{roubi1988resonance} alike mode. 
Despite the formal Raman-inactivity, both LO and TO components of infrared-active modes may be observed in RS spectrum due to long range Coulomb forces resulting from charges localised on atoms \cite{roubi1988resonance}. 
A resonant nature of RS involving such phonons was previously reported in Ref. \citenum{iwasaki1982Raman}. 
Temperature-dependent RS measurements \cite{golasa2017resonance} or measurements with modulation of excitation light \cite{Molas2017} might help to verify the attribution of the mode.
For the sake of this work, we assume the latter attribution to A$_{2u}$(LO) \cite{ibanez2018high}.  

The assignment of the weak feature at 141~cm$^{-1}$, previously reported in Ref.~\citenum{ibanez2018high}, is also not clear.
Its attribution to a second-order difference processes was suggested, as such peaks can be observed in RS spectra of some TMDs \cite{chen1974, golasa2013}. 
The E$_{u}$(TO) was also suggested as the signature of the mode \cite{roubi1988resonance, neal2021Raman}.
The peak in this work will be referred to as $\omega_{1}$. 

The effect of hydrostatic pressure on RS spectra of bulk \HfS during compression and decompression can be appreciated in Fig.~\ref{fig:1}(a) and (b), respectively.
Let us first focus on the results obtained during compression process. 
The lineshape of the RS spectra does not significantly change up to 5.7 GPa. 
The observed mode experience blueshifts with increasing of the hydrostatic pressure. Moreover, a substantial enhancement of the A$_{1g}$ peak can be noticed in the spectrum at 5.7 GPa.

\begin{figure}[h]
	\centering
	\includegraphics[width=.45\textwidth]{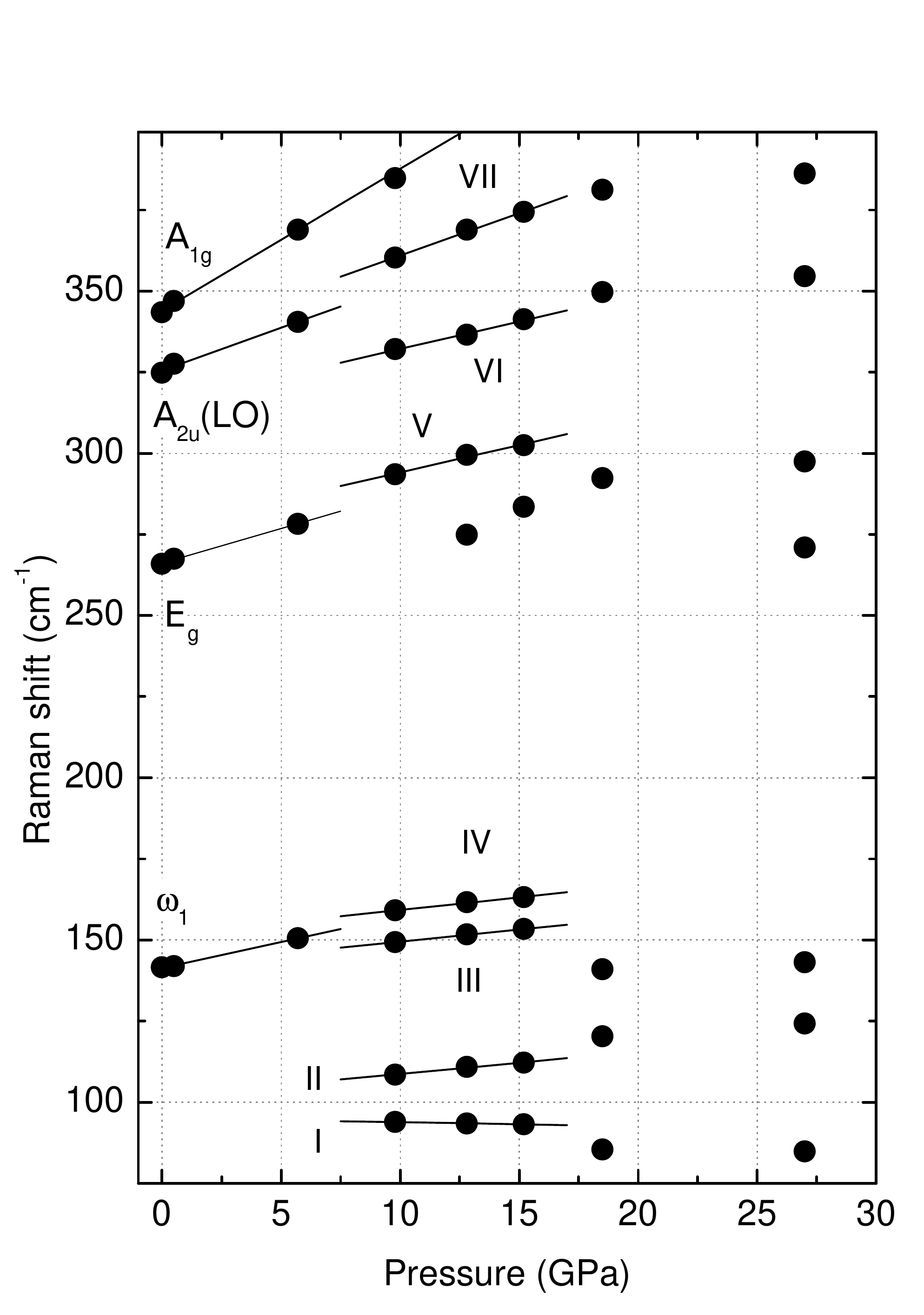}
\caption{The evolution of the Raman scattering peak energies in bulk \HfS with hydrostatic pressure. Straight lines correspond to linear fits to experimental data.}\label{fig:3}
\end{figure}

It was found that the energies of the observed modes increase linearly with increasing pressure, which can be appreciated in Fig.~\ref{fig:3}.
Assuming the linear dependence of the energies on pressure:\\
\vspace{-35pt}
\begin{center}
\begin{equation*}
     E(P)=E_0+\alpha P
\end{equation*}
\end{center}
\vspace{-5pt}
\noindent one can fit the experimental data with the results shown in Table \ref{tab:table}.

The obtained pressure coefficients are in general agreement with those recently reported by Ib{\'a}{\~n}ez and coworkers \cite{ibanez2018high}.
The initial values $E_0$ and pressure coefficients $\alpha$ for the observed modes span from 1.59~cm$^{-1}$/GPa to 4.36~cm$^{-1}$/GPa for A$_{1g}$ mode.

\begin{table}[t]
\caption{\label{tab:table} Results of linear fits to the pressure evolution of the Raman scattering peaks in bulk \HfS }
\begin{ruledtabular}
\begin{tabular}{c c c}
 Peak & $E_0$ &  $\alpha$ \\ 
  & [cm$^{-1}$] & [cm$^{-1}$/GPa] \\ \hline
  $\omega_1$ & 141.4 & 1.59 \\
 E$_g$ & 266.1 & 2.14 \\
 A$_{2u}$(LO) & 325.5 & 2.63\\
 A$_{1g}$ & 344.0 & 4.36 \\\hline
 I & 95.1 & -0.13 \\
 II & 101.8 & 0.69 \\
 III & 142.0 & 0.75 \\
 IV & 151.6 & 0.77 \\
 V & 277.3 & 1.68 \\
 VI & 315.3 & 1.69 \\
 VII & 334.9 & 2.61 \\
\end{tabular}
\end{ruledtabular}
\end{table}

\begin{figure}[b]
	\includegraphics[width=.5\textwidth]{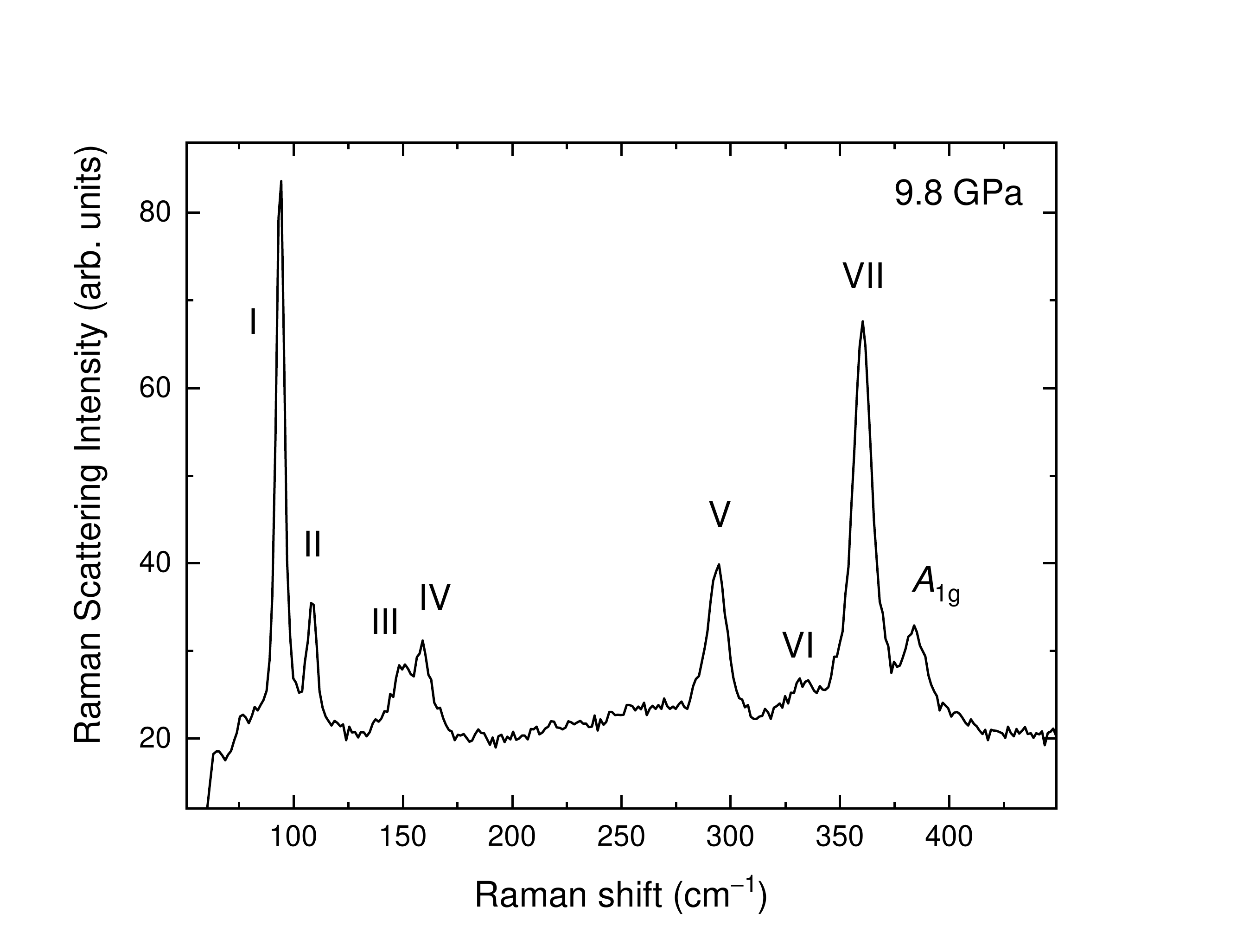}
\caption{Raman scattering spectrum of bulk \HfS in hydrostatic pressure of 9.8~GPa.} \label{fig:4}
\end{figure}

Substantial change in the line-shape of the RS spectrum can be observed at pressure of 9.8 GPa, which corresponding spectrum is presented in Fig.~\ref{fig:4}.
It can be appreciated in the Figure that the~A$_{1g}$ mode substantially weakens.
Simultaneously E$_g$, A$_{2u}$(LO), and $\omega_{1}$ disappear from the spectra and a set of new lines emerge in their stead. The~peaks are denoted with I-VII.
Similar lineshape of the spectra can be observed under hydrostatic pressure up to 12.8~GPa with the~A$_{1g}$ mode completely vanishing above 10~GPa.
The structural change of \HfS, which affects the RS spectra, was previously suggested in  Ref.~\citenum{ibanez2018high} with a new peak on the higher-frequency side of the A$_{1g}$ mode.
In our experiment, the transformation manifests itself in the emergence of peak VII at the lower-frequency side of the A$_{1g}$, as well as in changes to the low-energy part of the spectrum, in which two pairs of lines: I-II and III-IV can be observed.
The pressure coefficients $\alpha$ of the I-VII modes do not correspond in to those of modes detected at lower pressure.
In particular, the pressure coefficient $\alpha$ (2.61~cm$^{-1}$/GPa) for mode VII is substantially lower than the pressure coefficient for A$_{1g}$ mode although both linear energy evolutions have rather similar $E_0$ initial values  (334.9~cm$^{-1}$ for VII mode vs 344.0~cm$^{-1}$ for A$_{1g}$).
Similar effect can observed for E$_g$ and V mode with $E_0$ and $\alpha$ equal respectively to 266.1~cm$^{-1}$ and vs 277.3~cm$^{-1}$ and 2.14~cm$^{-1}$/GPa vs 1.68~cm$^{-1}$/GPa (see Table \ref{tab:table}).
Modes II-IV share relatively small pressure coefficient $\alpha$ equal to approx. 0.7~cm$^{-1}$/GPa) which is significantly smaller than the lowest coefficient for modes below the pressure-induced transformation.
Finally, the pressure coefficient for the I mode is negative. The properties substantially differ from those observed at lower pressure one can conclude that the phase transition of yet unknown origin takes place between 5.7 GPa and 9.8 GPa.

Next transformation of the RS spectrum can be observed above 15.2 GPa (see Fig.~\ref{fig:1}(a)).
Instead of seven well-defined vibrational modes present in the RS spectrum at lower pressure, broad spectral bands can be detected at 15.2 GPa or higher pressures. 
The overall lineshape of the RS spectra at highest hydrostatic pressure resembles that of disordered materials \cite{golasa2015disorder, minguzzi2015disorder}. 
The disorder and/or defects affecting crystalline structure can localize phonons allowing for momentum non-conserving processes.
As a result, the inelastic scattering is possible with phonons from whole Brilloiun zone.
Therefore the RS spectrum of disordered materials reflect the total density of phonon modes.
Adopting such an approach one can conclude that the low-energy band (below 150~cm$^{-1}$) corresponds to acoustic phonons and the high-energy band (at approx. 300~cm$^{-1}$) is due to optical vibrations.
The emergence of the disorder-like RS under highest hydrostatic pressure may be related to the contribution of an undefined phase observed in X-ray diffraction measurements as it was mentioned earlier.
As can be appreciated in the Fig.~\ref{fig:1}(b), the lineshape of RS spectra does not change during decompression process. 
It suggests permanent nature of the high-pressure transition, most likely into an amorphous phase. 

\section{Conclusions \label{sec:Conclusions}}

We have studied Raman scattering in bulk \HfS as a function of hydrostatic pressure up to 27~GPa. There are two transformation of Raman scattering spectrum observed during compression between 5.7~GPa and 9.8~GPa as well as between 12.8~GPa and 15.2~GPa. 
The spectrum after the lower-pressure transformation consists of seven well-defined vibrational modes as compared to four modes observed below the transformation. 
The modification in the spectrum suggests structural change in the material.
Frequencies of the modes observed above the transformation change linearly with pressure and the corresponding pressure coefficients have been determined.
Eelatively small (II-IV) or negative pressure coefficients (I) of the vibrational modes in the low-energy part of the spectra have been noticed.
The high-pressure transformation of the Raman scattering spectrum results in the change of its lineshape.
While a series of well-defined vibrational modes are observed under pressure below the transition, broad spectral bands can be detected at higher pressure. 
The lineshape of the spectra resembles that of disordered materials, which suggests amorphisation process involved. 
The lineshape does not change during decompression confirming permanent nature of the high-pressure transition.  

\section*{Acknowledgment}
The work has been supported by the National Science Centre, Poland (grants no.  2017/27/B/ST3/00205 and 2018/31/B/ST3/02111). Z.M. and W.Z. acknowledge support from the National Natural Science Foundation of China (Grants No. 61627813 and 61971024), the International Collaboration Project (No. B16001), the Beihang Hefei Innovation Research Institute (project no. BHKX-19-02).

\bibliographystyle{apsrev4-1}
\bibliography{biblio}

\begin{thebibliography}{16}%
\makeatletter
\providecommand \@ifxundefined [1]{%
 \@ifx{#1\undefined}
}%
\providecommand \@ifnum [1]{%
 \ifnum #1\expandafter \@firstoftwo
 \else \expandafter \@secondoftwo
 \fi
}%
\providecommand \@ifx [1]{%
 \ifx #1\expandafter \@firstoftwo
 \else \expandafter \@secondoftwo
 \fi
}%
\providecommand \natexlab [1]{#1}%
\providecommand \enquote  [1]{``#1''}%
\providecommand \bibnamefont  [1]{#1}%
\providecommand \bibfnamefont [1]{#1}%
\providecommand \citenamefont [1]{#1}%
\providecommand \href@noop [0]{\@secondoftwo}%
\providecommand \href [0]{\begingroup \@sanitize@url \@href}%
\providecommand \@href[1]{\@@startlink{#1}\@@href}%
\providecommand \@@href[1]{\endgroup#1\@@endlink}%
\providecommand \@sanitize@url [0]{\catcode `\\12\catcode `\$12\catcode
  `\&12\catcode `\#12\catcode `\^12\catcode `\_12\catcode `\%12\relax}%
\providecommand \@@startlink[1]{}%
\providecommand \@@endlink[0]{}%
\providecommand \url  [0]{\begingroup\@sanitize@url \@url }%
\providecommand \@url [1]{\endgroup\@href {#1}{\urlprefix }}%
\providecommand \urlprefix  [0]{URL }%
\providecommand \Eprint [0]{\href }%
\providecommand \doibase [0]{http://dx.doi.org/}%
\providecommand \selectlanguage [0]{\@gobble}%
\providecommand \bibinfo  [0]{\@secondoftwo}%
\providecommand \bibfield  [0]{\@secondoftwo}%
\providecommand \translation [1]{[#1]}%
\providecommand \BibitemOpen [0]{}%
\providecommand \bibitemStop [0]{}%
\providecommand \bibitemNoStop [0]{.\EOS\space}%
\providecommand \EOS [0]{\spacefactor3000\relax}%
\providecommand \BibitemShut  [1]{\csname bibitem#1\endcsname}%
\let\auto@bib@innerbib\@empty
\bibitem [{\citenamefont {Wilson}\ and\ \citenamefont
  {Yoffe}(1969)}]{wilson1969}%
  \BibitemOpen
  \bibfield  {author} {\bibinfo {author} {\bibfnamefont {A.}~\bibnamefont
  {Wilson}}\ and\ \bibinfo {author} {\bibfnamefont {A.~D.}\ \bibnamefont
  {Yoffe}},\ }\href {\doibase 10.1080/00018736900101307} {\bibfield  {journal}
  {\bibinfo  {journal} {Advances in Physics}\ }\textbf {\bibinfo {volume}
  {18}},\ \bibinfo {pages} {193} (\bibinfo {year} {1969})}\BibitemShut
  {NoStop}%
\bibitem [{\citenamefont {Xu}\ \emph {et~al.}(2015)\citenamefont {Xu},
  \citenamefont {Wang}, \citenamefont {Wang}, \citenamefont {Huang},
  \citenamefont {Wang}, \citenamefont {Yin}, \citenamefont {Jiang},\ and\
  \citenamefont {He}}]{xu2015ultrasensitive}%
  \BibitemOpen
  \bibfield  {author} {\bibinfo {author} {\bibfnamefont {K.}~\bibnamefont
  {Xu}}, \bibinfo {author} {\bibfnamefont {Z.}~\bibnamefont {Wang}}, \bibinfo
  {author} {\bibfnamefont {F.}~\bibnamefont {Wang}}, \bibinfo {author}
  {\bibfnamefont {Y.}~\bibnamefont {Huang}}, \bibinfo {author} {\bibfnamefont
  {F.}~\bibnamefont {Wang}}, \bibinfo {author} {\bibfnamefont {L.}~\bibnamefont
  {Yin}}, \bibinfo {author} {\bibfnamefont {C.}~\bibnamefont {Jiang}}, \ and\
  \bibinfo {author} {\bibfnamefont {J.}~\bibnamefont {He}},\ }\href {\doibase
  10.1002/adma.201503864} {\bibfield  {journal} {\bibinfo  {journal} {Advanced
  Materials}\ }\textbf {\bibinfo {volume} {27}},\ \bibinfo {pages} {7881}
  (\bibinfo {year} {2015})}\BibitemShut {NoStop}%
\bibitem [{\citenamefont {Kanazawa}\ \emph {et~al.}(2016)\citenamefont
  {Kanazawa}, \citenamefont {Amemiya}, \citenamefont {Ishikawa}, \citenamefont
  {Upadhyaya}, \citenamefont {Tsuruta}, \citenamefont {Tanaka},\ and\
  \citenamefont {Miyamoto}}]{kanazawa2016transistor}%
  \BibitemOpen
  \bibfield  {author} {\bibinfo {author} {\bibfnamefont {T.}~\bibnamefont
  {Kanazawa}}, \bibinfo {author} {\bibfnamefont {T.}~\bibnamefont {Amemiya}},
  \bibinfo {author} {\bibfnamefont {A.}~\bibnamefont {Ishikawa}}, \bibinfo
  {author} {\bibfnamefont {V.}~\bibnamefont {Upadhyaya}}, \bibinfo {author}
  {\bibfnamefont {K.}~\bibnamefont {Tsuruta}}, \bibinfo {author} {\bibfnamefont
  {T.}~\bibnamefont {Tanaka}}, \ and\ \bibinfo {author} {\bibfnamefont
  {Y.}~\bibnamefont {Miyamoto}},\ }\href {\doibase 10.1038/srep22277}
  {\bibfield  {journal} {\bibinfo  {journal} {Scientific reports}\ }\textbf
  {\bibinfo {volume} {6}},\ \bibinfo {pages} {22277} (\bibinfo {year}
  {2016})}\BibitemShut {NoStop}%
\bibitem [{\citenamefont {Ib{\'a}{\~n}ez}\ \emph {et~al.}(2018)\citenamefont
  {Ib{\'a}{\~n}ez}, \citenamefont {Wo{\'z}niak}, \citenamefont {Dybala},
  \citenamefont {Oliva}, \citenamefont {Hern{\'a}ndez},\ and\ \citenamefont
  {Kudrawiec}}]{ibanez2018high}%
  \BibitemOpen
  \bibfield  {author} {\bibinfo {author} {\bibfnamefont {J.}~\bibnamefont
  {Ib{\'a}{\~n}ez}}, \bibinfo {author} {\bibfnamefont {T.}~\bibnamefont
  {Wo{\'z}niak}}, \bibinfo {author} {\bibfnamefont {F.}~\bibnamefont {Dybala}},
  \bibinfo {author} {\bibfnamefont {R.}~\bibnamefont {Oliva}}, \bibinfo
  {author} {\bibfnamefont {S.}~\bibnamefont {Hern{\'a}ndez}}, \ and\ \bibinfo
  {author} {\bibfnamefont {R.}~\bibnamefont {Kudrawiec}},\ }\href {\doibase
  10.1038/s41598-018-31051-y} {\bibfield  {journal} {\bibinfo  {journal}
  {Scientific reports}\ }\textbf {\bibinfo {volume} {8}},\ \bibinfo {pages} {1}
  (\bibinfo {year} {2018})}\BibitemShut {NoStop}%
\bibitem [{\citenamefont {Lucovsky}\ \emph {et~al.}(1973)\citenamefont
  {Lucovsky}, \citenamefont {White}, \citenamefont {Benda},\ and\ \citenamefont
  {Revelli}}]{lukovsky1973IR}%
  \BibitemOpen
  \bibfield  {author} {\bibinfo {author} {\bibfnamefont {G.}~\bibnamefont
  {Lucovsky}}, \bibinfo {author} {\bibfnamefont {R.~M.}\ \bibnamefont {White}},
  \bibinfo {author} {\bibfnamefont {J.~A.}\ \bibnamefont {Benda}}, \ and\
  \bibinfo {author} {\bibfnamefont {J.~F.}\ \bibnamefont {Revelli}},\ }\href
  {\doibase 10.1103/PhysRevB.7.3859} {\bibfield  {journal} {\bibinfo  {journal}
  {Phys. Rev. B}\ }\textbf {\bibinfo {volume} {7}},\ \bibinfo {pages} {3859}
  (\bibinfo {year} {1973})}\BibitemShut {NoStop}%
\bibitem [{\citenamefont {Ib{\'a}{\~n}ez}()}]{ibanezPrivate}%
  \BibitemOpen
  \bibfield  {author} {\bibinfo {author} {\bibfnamefont {J.}~\bibnamefont
  {Ib{\'a}{\~n}ez}},\ }\href@noop {} {\bibinfo  {journal} {private
  information}\ }\BibitemShut {NoStop}%
\bibitem [{\citenamefont {Neal}\ \emph {et~al.}(2021)\citenamefont {Neal},
  \citenamefont {Li}, \citenamefont {Birol},\ and\ \citenamefont
  {Musfeldt}}]{neal2021Raman}%
  \BibitemOpen
\bibfield  {journal} {  }\bibfield  {author} {\bibinfo {author} {\bibfnamefont
  {S.~N.}\ \bibnamefont {Neal}}, \bibinfo {author} {\bibfnamefont
  {S.}~\bibnamefont {Li}}, \bibinfo {author} {\bibfnamefont {T.~J.}\
  \bibnamefont {Birol}}, \ and\ \bibinfo {author} {\bibfnamefont
  {L.}~\bibnamefont {Musfeldt}},\ }\href {\doibase 10.1038/s41699-021-00226-z}
  {\bibfield  {journal} {\bibinfo  {journal} {npj 2D Mater Appl}\ }\textbf
  {\bibinfo {volume} {5}},\ \bibinfo {pages} {45} (\bibinfo {year}
  {2021})}\BibitemShut {NoStop}%
\bibitem [{\citenamefont {Roubi}\ and\ \citenamefont
  {Carlone}(1988)}]{roubi1988resonance}%
  \BibitemOpen
  \bibfield  {author} {\bibinfo {author} {\bibfnamefont {L.}~\bibnamefont
  {Roubi}}\ and\ \bibinfo {author} {\bibfnamefont {C.}~\bibnamefont
  {Carlone}},\ }\href {\doibase 10.1103/PhysRevB.37.6808} {\bibfield  {journal}
  {\bibinfo  {journal} {Physical Review B}\ }\textbf {\bibinfo {volume} {37}},\
  \bibinfo {pages} {6808} (\bibinfo {year} {1988})}\BibitemShut {NoStop}%
\bibitem [{\citenamefont {Cingolani}\ \emph {et~al.}(1987)\citenamefont
  {Cingolani}, \citenamefont {Lugara}, \citenamefont {Scamarcio},\ and\
  \citenamefont {L{\'e}vy}}]{cingolani1987raman}%
  \BibitemOpen
  \bibfield  {author} {\bibinfo {author} {\bibfnamefont {A.}~\bibnamefont
  {Cingolani}}, \bibinfo {author} {\bibfnamefont {M.}~\bibnamefont {Lugara}},
  \bibinfo {author} {\bibfnamefont {G.}~\bibnamefont {Scamarcio}}, \ and\
  \bibinfo {author} {\bibfnamefont {F.}~\bibnamefont {L{\'e}vy}},\ }\href
  {\doibase 10.1016/0038-1098(87)91126-4} {\bibfield  {journal} {\bibinfo
  {journal} {Solid State Communications}\ }\textbf {\bibinfo {volume} {62}},\
  \bibinfo {pages} {121} (\bibinfo {year} {1987})}\BibitemShut {NoStop}%
\bibitem [{\citenamefont {Iwasaki}\ \emph {et~al.}(1982)\citenamefont
  {Iwasaki}, \citenamefont {Kuroda},\ and\ \citenamefont
  {Y}}]{iwasaki1982Raman}%
  \BibitemOpen
  \bibfield  {author} {\bibinfo {author} {\bibfnamefont {T.}~\bibnamefont
  {Iwasaki}}, \bibinfo {author} {\bibfnamefont {N.}~\bibnamefont {Kuroda}}, \
  and\ \bibinfo {author} {\bibfnamefont {N.}~\bibnamefont {Y}},\ }\href
  {\doibase 10.1143/JPSJ.51.2233} {\bibfield  {journal} {\bibinfo  {journal}
  {J. Phys. Soc. Jpn.}\ }\textbf {\bibinfo {volume} {52}},\ \bibinfo {pages}
  {2233} (\bibinfo {year} {1982})}\BibitemShut {NoStop}%
\bibitem [{\citenamefont {Go\l{}asa}\ \emph {et~al.}(2017)\citenamefont
  {Go\l{}asa}, \citenamefont {Grzeszczyk}, \citenamefont {Molas}, \citenamefont
  {Zinkiewicz}, \citenamefont {Bala}, \citenamefont {Nogajewski}, \citenamefont
  {Potemski}, \citenamefont {Wysmo\l{}ek},\ and\ \citenamefont
  {Babi\'nski}}]{golasa2017resonance}%
  \BibitemOpen
  \bibfield  {author} {\bibinfo {author} {\bibfnamefont {K.}~\bibnamefont
  {Go\l{}asa}}, \bibinfo {author} {\bibfnamefont {M.}~\bibnamefont
  {Grzeszczyk}}, \bibinfo {author} {\bibfnamefont {M.~R.}\ \bibnamefont
  {Molas}}, \bibinfo {author} {\bibfnamefont {M.}~\bibnamefont {Zinkiewicz}},
  \bibinfo {author} {\bibfnamefont {L.}~\bibnamefont {Bala}}, \bibinfo {author}
  {\bibfnamefont {K.}~\bibnamefont {Nogajewski}}, \bibinfo {author}
  {\bibfnamefont {M.}~\bibnamefont {Potemski}}, \bibinfo {author}
  {\bibfnamefont {A.}~\bibnamefont {Wysmo\l{}ek}}, \ and\ \bibinfo {author}
  {\bibfnamefont {A.}~\bibnamefont {Babi\'nski}},\ }\href {\doibase
  10.1515/nanoph-2016-0150} {\bibfield  {journal} {\bibinfo  {journal}
  {Nanophotonics}\ }\textbf {\bibinfo {volume} {6}},\ \bibinfo {pages} {1281}
  (\bibinfo {year} {2017})}\BibitemShut {NoStop}%
\bibitem [{\citenamefont {Molas}\ \emph {et~al.}(2017)\citenamefont {Molas},
  \citenamefont {Nogajewski}, \citenamefont {Potemski},\ and\ \citenamefont
  {Babi{\'{n}}ski}}]{Molas2017}%
  \BibitemOpen
  \bibfield  {author} {\bibinfo {author} {\bibfnamefont {M.~R.}\ \bibnamefont
  {Molas}}, \bibinfo {author} {\bibfnamefont {K.}~\bibnamefont {Nogajewski}},
  \bibinfo {author} {\bibfnamefont {M.}~\bibnamefont {Potemski}}, \ and\
  \bibinfo {author} {\bibfnamefont {A.}~\bibnamefont {Babi{\'{n}}ski}},\ }\href
  {\doibase 10.1038/s41598-017-05367-0} {\bibfield  {journal} {\bibinfo
  {journal} {Scientific Reports}\ }\textbf {\bibinfo {volume} {7}},\ \bibinfo
  {pages} {5036} (\bibinfo {year} {2017})}\BibitemShut {NoStop}%
\bibitem [{\citenamefont {Chen}\ and\ \citenamefont {Wang}(1974)}]{chen1974}%
  \BibitemOpen
  \bibfield  {author} {\bibinfo {author} {\bibfnamefont {J.~M.}\ \bibnamefont
  {Chen}}\ and\ \bibinfo {author} {\bibfnamefont {C.~S.}\ \bibnamefont
  {Wang}},\ }\href {\doibase /10.1016/0038-1098(74)90150-1} {\bibfield
  {journal} {\bibinfo  {journal} {Solid State Communications}\ }\textbf
  {\bibinfo {volume} {14}},\ \bibinfo {pages} {857} (\bibinfo {year}
  {1974})}\BibitemShut {NoStop}%
\bibitem [{\citenamefont {Go\l{}asa}\ \emph {et~al.}(2013)\citenamefont
  {Go\l{}asa}, \citenamefont {Grzeszczyk}, \citenamefont {Korona},
  \citenamefont {Bo\.zek}, \citenamefont {Binder}, \citenamefont {Szczytko},
  \citenamefont {Wysmo\l{}ek},\ and\ \citenamefont {Babi\'nski}}]{golasa2013}%
  \BibitemOpen
  \bibfield  {author} {\bibinfo {author} {\bibfnamefont {K.}~\bibnamefont
  {Go\l{}asa}}, \bibinfo {author} {\bibfnamefont {M.}~\bibnamefont
  {Grzeszczyk}}, \bibinfo {author} {\bibfnamefont {K.~P.}\ \bibnamefont
  {Korona}}, \bibinfo {author} {\bibfnamefont {R.}~\bibnamefont {Bo\.zek}},
  \bibinfo {author} {\bibfnamefont {J.}~\bibnamefont {Binder}}, \bibinfo
  {author} {\bibfnamefont {J.}~\bibnamefont {Szczytko}}, \bibinfo {author}
  {\bibfnamefont {A.}~\bibnamefont {Wysmo\l{}ek}}, \ and\ \bibinfo {author}
  {\bibfnamefont {A.}~\bibnamefont {Babi\'nski}},\ }\href {\doibase
  10.12693/APhysPolA.124.849} {\bibfield  {journal} {\bibinfo  {journal} {Acta
  Physica Polonica A}\ }\textbf {\bibinfo {volume} {124}},\ \bibinfo {pages}
  {849} (\bibinfo {year} {2013})}\BibitemShut {NoStop}%
\bibitem [{\citenamefont {Go\l{}asa}\ \emph {et~al.}(2015)\citenamefont
  {Go\l{}asa}, \citenamefont {Grzeszczyk}, \citenamefont {Binder},
  \citenamefont {Bo\.zek}, \citenamefont {Wysmo\l{}ek},\ and\ \citenamefont
  {Babi\'nski}}]{golasa2015disorder}%
  \BibitemOpen
  \bibfield  {author} {\bibinfo {author} {\bibfnamefont {K.}~\bibnamefont
  {Go\l{}asa}}, \bibinfo {author} {\bibfnamefont {M.}~\bibnamefont
  {Grzeszczyk}}, \bibinfo {author} {\bibfnamefont {J.}~\bibnamefont {Binder}},
  \bibinfo {author} {\bibfnamefont {R.}~\bibnamefont {Bo\.zek}}, \bibinfo
  {author} {\bibfnamefont {A.}~\bibnamefont {Wysmo\l{}ek}}, \ and\ \bibinfo
  {author} {\bibfnamefont {A.}~\bibnamefont {Babi\'nski}},\ }\href {\doibase
  10.1063/1.4926670} {\bibfield  {journal} {\bibinfo  {journal} {AIP Advances}\
  }\textbf {\bibinfo {volume} {5}},\ \bibinfo {pages} {077120} (\bibinfo {year}
  {2015})}\BibitemShut {NoStop}%
\bibitem [{\citenamefont {Mignuzzi}\ \emph {et~al.}(2015)\citenamefont
  {Mignuzzi}, \citenamefont {Pollard}, \citenamefont {Bonini}, \citenamefont
  {Brennan}, \citenamefont {Gilmore}, \citenamefont {Pimenta}, \citenamefont
  {Richards},\ and\ \citenamefont {Roy}}]{minguzzi2015disorder}%
  \BibitemOpen
  \bibfield  {author} {\bibinfo {author} {\bibfnamefont {S.}~\bibnamefont
  {Mignuzzi}}, \bibinfo {author} {\bibfnamefont {A.~J.}\ \bibnamefont
  {Pollard}}, \bibinfo {author} {\bibfnamefont {N.}~\bibnamefont {Bonini}},
  \bibinfo {author} {\bibfnamefont {B.}~\bibnamefont {Brennan}}, \bibinfo
  {author} {\bibfnamefont {I.~S.}\ \bibnamefont {Gilmore}}, \bibinfo {author}
  {\bibfnamefont {M.~A.}\ \bibnamefont {Pimenta}}, \bibinfo {author}
  {\bibfnamefont {D.}~\bibnamefont {Richards}}, \ and\ \bibinfo {author}
  {\bibfnamefont {D.}~\bibnamefont {Roy}},\ }\href {\doibase
  10.1103/PhysRevB.91.195411} {\bibfield  {journal} {\bibinfo  {journal} {Phys.
  Rev. B}\ }\textbf {\bibinfo {volume} {91}},\ \bibinfo {pages} {195411}
  (\bibinfo {year} {2015})}\BibitemShut {NoStop}%
\end{thebibliography}%
\end{document}